\documentclass[twocolumn,showpacs,preprintnumbers,amsmath,amssymb,pre,nofootinbib,superscriptaddress]{revtex4}
\usepackage{psfrag}
\usepackage{graphicx}
\usepackage{bm}
\newcommand{\ve}{\varepsilon}

\newcommand{\brat}[1]{\ensuremath{\left<#1\right>}}

\newcommand{\braAket}[3]{\ensuremath{\left<#1\vphantom{#2}\vphantom{#3}
\right|#2\left|#3\vphantom{#1}\vphantom{#2}\right>}}

\begin{document}

\title{Scaling approach to related disordered stochastic and free-fermion models}
\author{R.~J. Harris}
\altaffiliation{Present address: Fachrichtung Theoretische Physik, Universit\"at des Saarlandes, 66041 Saarbr\"ucken, Germany}
\email{harris@lusi.uni-sb.de}
\affiliation{Institut f\"ur Festk\"orperforschung, Forschungszentrum J\"ulich, 52425 J\"ulich, Germany}
\author{R.~B. Stinchcombe}
\email{stinch@thphys.ox.ac.uk}
\affiliation{Rudolf Peierls Centre for Theoretical Physics, University of Oxford, 1 Keble Road, Oxford, OX1 3NP,UK}

\date{January 2, 2007}

\begin{abstract}
Motivated by mapping from a stochastic system with spatially random rates, we consider disordered non-conserving free-fermion systems using a scaling procedure for the equations of motion.  This approach demonstrates disorder-induced localization acting in competition with the asymmetric driving.  We discuss the resulting implications for the original stochastic system.
\end{abstract}


\maketitle

\section{Introduction}
\label{s:ffintro}

The role of disorder in stochastic non-equilibrium systems is a topic of much recent interest (see e.g.,~\cite{Schmittmann95,Tripathy98,Laessing98,Stinchcombe02,Evans04d} and references therein).   
This is exemplified by studies of the asymmetric simple exclusion process (ASEP), a one-dimensional lattice gas model exhibiting a phase transition for open boundary conditions.  The effect of quenched particle-disordered hop rates in this system was investigated in, e.g.,~\cite{Krug96,Evans96,Jain03} and the case of quenched spatial disorder has been clarified by a number of works (see, e.g.,~\cite{Tripathy98, Krug00,Kolwankar00, Juhasz05b}) including via a linearization transformation and scaling approach~\cite{Me04}.

In this paper we study a related non-equilibrium model with applications in catalysis---the partially asymmetric exclusion process combined with deposition and evaporation of dimers (or equivalently pair annihilation and creation).  Studies of the pure model e.g.,~\cite{Grynberg95,Grynberg96,Santos97} suggest that, unlike the ASEP, for adsorption and desorption rates which do not vanish in the thermodynamic limit, there is no steady-state phase transition even with open boundary conditions~\cite{Alcaraz93, Peschel94}.  However, the effect of disorder on both steady-state properties (such as the density) and dynamics is still of interest.

Here we demonstrate that, for spatially disordered rates obeying a particular condition, one can approach this problem via a powerful mapping to free-fermion systems.  In the pure case, the resulting free-fermion system can be treated by Fourier and Bogoliubov transformations~\cite{Grynberg94,Grynberg95}.  The simplest disordered scenario, the case of a single defect, is known to produce the stochastic analogue of localized modes~\cite{Chen94}.  One might also expect to make progress for the case of dilution, by breaking the chain up into finite uniform sections (compare the treatment of the diffusion-only problem in~\cite{Grynberg00}; a similar break-up into effective pure segments also occurs in~\cite{Evans04d}).  In this paper we are chiefly interested in more general disordered cases where there is no translational invariance and therefore one can not utilize the usual Fourier and Bogoliubov transformations.  However, work by Merz and Chalker involving doubling the number of fermion operators~\cite{Merz02}, offers the possibility of using a type of a localized Bogoliubov transformation followed by numerical calculation.  Another potential way to make progress is to derive and work with the linear dynamic equations and this is the approach pursued in the present work.

Our scaling approach is based on a method developed by Pimentel and Stinchcombe~\cite{Pimentel88} to treat the equation of motion of a 1D Mattis-transformed Edwards-Anderson Heisenberg spin glass and later applied by the present authors to the Cole-Hopf transformed ASEP~\cite{Me04}.  Essentially, one looks at the evolution of parameters after a $b=2$ decimation of the equation of motion.  For extended states the parameters are found to evolve chaotically under the scaling whereas for localized states they decay exponentially allowing the identification of a localization length.  Application of this method to the present problem contributes to understanding of disordered free-fermions as well as illustrating a potential way to make progress with the disordered stochastic model.

The plan of the rest of the paper is as follows.  In Sec.~\ref{s:quantum} we define our model, introduce the quantum Hamiltonian formalism and demonstrate the free-fermion reduction for a particular choice of disorder.  We also discuss the known pure results.  In Sec.~\ref{s:motion}, we derive the coupled equations of motions for the disordered free-fermion case.  Then, in Sec.~\ref{s:scale} we exploit the fact that these equations are linear to apply scaling techniques based on those in~\cite{Pimentel88,Me04}.  As in those works, we find that disorder induces localization effects and in Sec.~\ref{s:origsto} we tentatively discuss the consequences for the original stochastic system.  Finally, Sec.~\ref{s:ffdiss} contains conclusions and suggestions for future work.

\section{Model and mappings}
\label{s:quantum}

\begin{figure}
\begin{center}
\psfrag{pl}[Bc][Tc]{$p$}
\psfrag{ql}[Bc][Tc]{$q$}
\psfrag{e}[Cl][Cl]{$\varepsilon$}
\psfrag{f}[Cl][Cl]{$\varepsilon'$}
\includegraphics*[width=0.8\columnwidth]{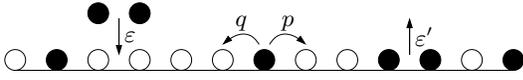}
\caption{Partially asymmetric exclusion process with dimer evaporation and deposition (pure model).  Filled circles indicate particles; open circles denote vacancies.}
\label{f:ffpure}
\end{center}
\end{figure}

Let us start by considering the pure partially asymmetric exclusion process with dimer evaporation and deposition, shown schematically in Fig.~\ref{f:ffpure}.
Subject to the exclusion constraints, particles hop to the right (left) with rate $p$ ($q$), and pairs of particles are deposited (evaporated) with rate $\varepsilon$ ($\varepsilon'$).  This provides a simple model for adsorption-desorption processes and catalytic surface reactions.

In the special case $p+q=\varepsilon+\varepsilon'$, the system can be mapped via a Jordan-Wigner transformation~\cite{Jordan28} and a Bogoliubov-Valatin transformation~\cite{Bogoliubov58,Valatin58} to a free-fermion problem and hence solved exactly~\cite{Alcaraz94,Grynberg94,Grynberg95,Grynberg96}.  The results are of more general significance since, by duality transformations and spin rotations, the quantum Hamiltonian can also be mapped to other stochastic systems~\cite{Henkel95,Santos97}, including some which are experimentally realizable~\cite{Kroon93}.  A subcase is $p=q=\varepsilon=\varepsilon'$ which maps to the Ising model and so is well understood~\cite{Grynberg94}.  

The aim of the present work is to study the spatially-disordered version of this model,with rates $p_l$, $q_l$, $\ve_l$ and $\ve'_l$ for processes on the bond between sites $l$ and $l+1$.  In this section we demonstrate that the mapping to free-fermions is still valid in the case where $p_l+q_l=\varepsilon_l+\varepsilon'_l$ for each $l$.

The generalized pair process parameterized by $p_l$, $q_l$, $\ve_l$ and $\ve'_l$ can be conveniently represented using the quantum Hamiltonian formalism~\cite{Schutz01}.  In this approach one defines a probability vector $|P\rangle = \sum_n P_n |n\rangle$ with $|n\rangle$ the basis vector associated with the particle configuration $n=(n_1,n_2,\ldots,n_L)$ and $P_n$ the probability measure on the set of all such configurations. $|P\rangle$ obeys the normalization condition $\langle s|P \rangle =1$ where $\langle s | = \sum_n \langle n |$ and $\langle n | n' \rangle = \delta_{n,n'}$.  Within this formalism the master equation for the time evolution resembles a Schr\"odinger equation
\begin{equation}
\frac{d}{dt}|P(t)\rangle = - H |P(t)\rangle \label{e:sch}
\end{equation}
with
\begin{multline}
H = \\
 -\sum_l \bigl[ p_l \sigma^-_l \sigma^+_{l+1} +  q_l \sigma^+_l \sigma^-_{l+1} + \ve_l \sigma^+_l \sigma^+_{l+1} + \ve'_l \sigma^-_l \sigma^-_{l+1} \\
 -\frac{1}{4}(p_l + q_l + \ve_l + \ve'_l) + \frac{1}{4}(-p_l + q_l + \ve_l -\ve'_l)\sigma^z_l \\
 + \frac{1}{4}(p_l - q_l + \ve_l - \ve'_l)\sigma^z_{l+1} + \frac{1}{4}(p_l + q_l - \ve_l - \ve'_l)\sigma^z_l \sigma^z_{l+1} \bigr]. \label{e:Hpair}
\end{multline}
Here the ``spin-flip'' operators $\sigma^\pm$ correspond to particle creation and annihilation processes while the projection operators $\sigma_z$ are needed to account for the probability that configurations do not change.  Conservation of probability imposes the condition
\begin{equation}
\langle s | H = 0. \label{e:probcon}
\end{equation}

Note that the general four-parameter model defined by~\eqref{e:Hpair} contains many other important models as subcases, e.g., the ASEP ($q_l=\ve_l=\ve'_l=0$), symmetric diffusion ($p_l=q_l$, $\ve_l=\ve'_l=0$), etc. Further progress depends on what boundary terms are imposed and how tractable the resultant quantum spin model is.  In particular, a Jordan-Wigner transformation to fermion operators $c_l^\dagger$, $c_l$ gives both an easily treatable quadratic part (arising from $\sigma^\pm \sigma^\pm$ and $\sigma^z$ terms) and a more difficult quartic part (from the $\sigma^z \sigma^z$ terms).  The quartic terms will clearly be zero for the special case 
\begin{equation}
p_l+q_l=\ve_l+\ve_l' \label{e:ff}
\end{equation} 
and this gives the disordered free-fermion model which will be explored in more detail in the remainder of this paper.

For this particular case, we now outline the details of the transformation to fermion operators (compare Grynberg \emph{et al.}~\cite{Grynberg94} for the pure model).  Writing $\sigma^z$ in terms of $\sigma^+ \sigma^-$ and imposing periodic boundary conditions,
Eq.~\eqref{e:Hpair} becomes
\begin{multline}
H_{\text{FF}} = -\sum_l [ p_l \sigma^-_l \sigma^+_{l+1} +  q_l \sigma^+_l \sigma^-_{l+1} + \ve_l \sigma^+_l \sigma^+_{l+1} + \ve'_l \sigma^-_l \sigma^-_{l+1} \\
+ (p_{l-1} - p_l +\ve_l - \ve'_{l-1}) \sigma^+_l \sigma^-_l - \ve_l ].
\end{multline}
Then applying the usual Jordan-Wigner transformation~\cite{Jordan28}
\begin{align}
\sigma_l^+ &= c_l^\dagger \exp \left( i \pi \sum_{j<l} c_j^\dagger c_j \right)\\
\sigma_l^- &= \exp \left( -i \pi \sum_{j<l} c_j^\dagger c_j \right) c_l,
\end{align}
yields the Hamiltonian in terms of fermion operators\footnote{If the total number of fermions, $\sum_l \sigma_l^+ \sigma_l^- = \sum_l c_l^\dagger c_l$, is even then the fermion operators must obey cyclic boundary conditions; in the odd subspace, anti-cyclic boundary conditions are required.  See~\cite{Grynberg94}.}
\begin{multline}
H_{\text{FF}}=\sum_l [ p_l c_l c^\dagger_{l+1} - q_l c_l^\dagger c_{l+1} - \ve_l (c_l^\dagger c^\dagger_{l+1} - 1)  \\
+ \ve'_l c_l c_{l+1} - (p_{l-1} - p_l +\ve_l - \ve'_{l-1})c_l^\dagger c_l ]. \label{e:HFF}
\end{multline}
The study of the disordered free-fermion Hamiltonian~\eqref{e:HFF} is the central aim of this paper; we now digress to summarize the known pure results.

In the pure case one can exploit translational invariance via a Fourier transformation and then use a Bogoliubov-type similarity transformation~\cite{Bogoliubov58,Valatin58} (well defined only for $\ve$ and $\ve'$ non-zero) to diagonalize the Hamiltonian.  The 
details are given in~\cite{Grynberg94,Grynberg95}; here we simply quote the resulting Hamiltonian:
\begin{equation}
H_{\text{FF}}=\sum_k \lambda_k \xi^+_k \xi_k,
\end{equation}
with spectrum 
\begin{equation}
\lambda_k = \ve + \ve' + (\ve - \ve') \cos k + i(p-q)\sin k. \label{e:spectrum}
\end{equation}
The imaginary part of $\lambda_k$ implies ballistic motion while the real part indicates that excitations decay with time constant $[\ve + \ve' + (\ve - \ve') \cos k]^{-1}$.  For $\ve$ and $\ve'$ non-zero, the spectrum is gapped giving exponentially fast kinetics.  By mapping back from this free-fermion model, 
one can obtain the steady-state density profile for the original pure stochastic problem.  The result~\cite{Grynberg94}
\begin{equation}
\rho_l=\frac{1}{1+\sqrt{\varepsilon'/\varepsilon}}, \label{e:dens}
\end{equation}
is in agreement with mean-field calculation (one can show that the system is spatially uncorrelated).  Other works have calculated dynamic correlation functions~\cite{Grynberg95,Grynberg96}, shock evolution~\cite{Santos96}, persistence probability, etc.  Note that non-steady-state properties of the stochastic system can be understood by mapping to the Glauber model~\cite{Glauber63} and treating the asymmetric hopping as biased diffusion of domain walls~\cite{Family91}.

In the disordered case where there is no translational invariance a different approach is required---in the next section, we demonstrate how to derive the equations of motion starting from the free-fermion Hamiltonian~\eqref{e:HFF}.

\section{Equations of motion}
\label{s:motion}

Let us first consider the expectation value of a general observable $X$, i.e.,
\begin{equation}
\brat{X}(t) = \braAket{s}{X}{P}. \label{e:Xav2}
\end{equation}
where we use angular brackets to denote this stochastic average over histories. 
Using the ``Schr{\"o}dinger equation''~\eqref{e:sch} one sees that the time evolution of $\brat{X}$ is given by
\begin{equation}
\frac{\partial\brat{X}}{\partial t} = \brat{[H,X]}
\end{equation}
with $[H,X]$ the usual commutator.

For the free-fermion Hamiltonian~\eqref{e:HFF} we can use the fermion anti-commutation relations, together with the identity $[A,BC]=\{A,B\}C - B\{A,C\}$, to calculate the commutators $[c_j,H]$ and $[c_j^\dagger,H]$ and hence obtain coupled dynamic equations for the operators (in the Heisenberg representation):
\begin{align}
\frac{\partial c_j}{\partial t} &= p_{j-1} c_{j-1} + q_j c_{j+1} -\ve_{j-1} c^\dagger_{j-1} +\ve_j c^\dagger_{j+1} \notag \\
& \phantom{=} +(p_{j-1} - p_j +\ve_j - \ve'_{j-1})c_j \label{e:cj}, \\
\frac{\partial c^\dagger_j} {\partial t} &= -p_j c^\dagger_{j+1} - q_{j-1} c^\dagger_{j-1} +\ve'_{j-1} c_{j-1} -\ve'_j c_{j+1} \notag \\
& \phantom{=} -(p_{j-1} - p_j +\ve_j - \ve'_{j-1}) c^\dagger_j. \label{e:cj+}
\end{align}
Note that due to the non-Hermicity of $H$ these equations are not simply Hermitian conjugates of each other.  The equations are linear thus raising the possibility of using scaling techniques like those in~\cite{Pimentel88,Me04}---this will be the subject of the next section. 

As a preliminary to this scaling approach,  we first assume that $c_j$ and $c^\dagger_j$ can be written as superpositions of operators with exponential time dependence:
\begin{align}
c_j&=\sum_\omega \Gamma_j(\omega) e^{\omega t}, \\
c^\dagger_j&=\sum_\omega \gamma_j(\omega) e^{\omega t}.
\end{align}
Substituting into Eqs.~\eqref{e:cj} and~\eqref{e:cj+} and equating components with the same time dependence yields
\begin{align}
(E_{j} - \omega)\Gamma_j &= -p_{j-1} \Gamma_{j-1} - q_j \Gamma_{j+1} + \ve_{j-1} \gamma_{j-1} -\ve_j \gamma_{j+1}, \label{e:alpha} \\
(E_{j} + \omega)\gamma_j &= -p_j \gamma_{j+1} - q_{j-1} \gamma_{j-1} +\ve'_{j-1} \Gamma_{j-1} -\ve'_j \Gamma_{j+1}, \label{e:gamma}
\end{align}
with $E_{j}=(p_{j-1} - p_j +\ve_j - \ve'_{j-1})$.  

Note that for the pure case we can carry out a Fourier transformation by writing $\Gamma_j(\omega)=e^{-ikj}\Gamma_k$, $\gamma_j(\omega)=e^{-ikj}\gamma_k$.  As expected, we then find that for each Fourier component, Eqs.~\eqref{e:alpha} and~\eqref{e:gamma} have solutions with $\omega=\lambda_k$ or $\omega=-\lambda_{-k}$, where $\lambda_k$ is the pure spectrum~\eqref{e:spectrum}.
In other words the operators $c_j$ and $c_j^\dagger$ can be written as
\begin{align}
c_j &= \sum_k e^{-ikj}(\Gamma_k e^{\lambda_k t} + \Gamma'_k e^{-\lambda_{-k} t}), \label{e:fbog1} \\
c_j^\dagger &= \sum_k e^{-ikj}(\gamma_k e^{\lambda_k t} + \gamma'_k e^{-\lambda_{-k} t}), \label{e:fbog2}
\end{align}
where
\begin{align}
{\gamma_k} &= i \Gamma_k \frac{\ve'}{\ve} \cot \frac{k}{2}, \label{e:gamrat1} \\
{\gamma'_k} &= -i \Gamma'_k \tan \frac{k}{2}. \label{e:gamrat2}
\end{align}
The Bogoliubov angle can be obtained from the relations~\eqref{e:gamrat1} and~\eqref{e:gamrat2}; in the pure case this approach thus provides an alternative route to direct transformation of the Hamiltonian~\eqref{e:HFF}.

As an aside, we remark that Eqs.~\eqref{e:alpha} and~\eqref{e:gamma} can also be cast in the form of a transfer matrix mapping.  Analysing products of disordered transfer matrices then provides a way to investigate possible localization effects.  This is a particularly attractive approach for simple models such as binary disorder.  In the next section we develop instead, a general numerical scaling approach for arbitrary distributions of disorder.

\section{Scaling}
\label{s:scale}

\subsection{Procedure}

In the general disordered case it is not obvious how to solve Eqs.~\eqref{e:alpha} and~\eqref{e:gamma} explicitly but their form is reminiscent of an equation which appears in the disordered ASEP (in a mean-field description) after a linearizing (Cole-Hopf) transformation so we can develop a scaling method similar to the one used in that case~\cite{Me04} (compare also the original work of Pimentel and Stinchcombe~\cite{Pimentel88}).  For full generality we must allow each coefficient to scale differently so we perform a $b=2$ dilation of the system using decimation on the following equations
\begin{align}
(E_{j} - \omega)\Gamma_j =& -p_{j,j-1} \Gamma_{j-1} - q_{j,j+1} \Gamma_{j+1} + \ve_{j,j-1} \gamma_{j-1} \notag \\
&-\tilde{\ve}_{j,j+1} \gamma_{j+1} \label{e:alpha2}, \\
(\tilde{E}_{j} + \omega)\gamma_j =& -\tilde{p}_{j,j+1} \gamma_{j+1} - \tilde{q}_{j,j-1} \gamma_{j-1} +\ve'_{j,j-1} \Gamma_{j-1} \notag \\
&-\tilde{\ve}'_{j,j+1} \Gamma_{j+1}. \label{e:gamma2}
\end{align}
where we have extended the notation for clarity.  For example, $p_{j,j-1}$ denotes the coefficient of $\Gamma_{j-1}$ in the expression for $\Gamma_j$.  Even in the pure case, this may scale differently from $\tilde{p}_{j,j+1}$ which is the coefficent of $\gamma_{j,j+1}$ in the expression for $\gamma_j$.  In the rescaled equation for $\Gamma_j$ ($\gamma_j$) the coefficient of $\Gamma_{j-2}$ ($\gamma_{j+2}$) is $p_{j,j-2}$ ($\tilde{p}_{j,j+2}$).  The resulting 10-parameter scaling equations are rather involved and are displayed for convenience in Appendix~\ref{a:ffscale}.

We now test this scaling procedure on the pure model (Sec.~\ref{ss:ffpure}) before applying it to the disordered problem (Sec.~\ref{ss:ffdis}).

\subsection{Pure model}
\label{ss:ffpure}

We here demonstrate how the scaling of the equations of motion~\eqref{e:alpha2} and~\eqref{e:gamma2} reflects the known spectrum $\lambda_k$ of the pure model~\eqref{e:spectrum}.  For definiteness we consider the case $\ve>\ve'$, $p \geq q$ throughout; the extension to other cases should be straightforward, since they are related by duality or space reflection.

\subsubsection{Non-biased case, $p = q$}

For the pure case, if one starts with $p=\tilde{p}=q=\tilde{q}$, $\ve=\tilde{\ve}$, $\ve'=\tilde{\ve}'$, and by implication $E=\tilde{E}$, then the relationships $p=q$, $\tilde{p}=\tilde{q}$, $\ve=\tilde{\ve}$ and $\ve'=\tilde{\ve}'$ are preserved under scaling so one only needs to scale six independent parameters. We find two distinct types of scaling behaviour.  

\begin{figure}
\begin{center}
\psfrag{I}[Tc][Bc]{$I$}
\psfrag{r}[Bc][Tc]{$\text{rescaled parameters}$}
\psfrag{0}[Cr][Cr]{\scriptsize{0.0}}
\psfrag{0.5}[Cr][Cr]{\scriptsize{0.5}}
\psfrag{1}[Cr][Cr]{\scriptsize{1.0}}
\psfrag{1.5}[Cr][Cr]{\scriptsize{1.5}}
\psfrag{2}[Cr][Cr]{\scriptsize{2.0}}
\psfrag{2.5}[Cr][Cr]{\scriptsize{2.5}}
\psfrag{3}[Cr][Cr]{\scriptsize{3.0}}
\psfrag{0b}[Tc][Tc]{\scriptsize{0}}
\psfrag{2b}[Tc][Tc]{\scriptsize{2}}
\psfrag{4}[Tc][Tc]{\scriptsize{4}}
\psfrag{6}[Tc][Tc]{\scriptsize{6}}
\psfrag{8}[Tc][Tc]{\scriptsize{8}}
\psfrag{10}[Tc][Tc]{\scriptsize{10}}
\psfrag{12}[Tc][Tc]{\scriptsize{12}}
\psfrag{14}[Tc][Tc]{\scriptsize{14}}
\psfrag{16}[Tc][Tc]{\scriptsize{16}}
\psfrag{18}[Tc][Tc]{\scriptsize{18}}
\psfrag{20}[Tc][Tc]{\scriptsize{20}}
\psfrag{"inbp1.dat"}[Cr][Cr]{\tiny{$p$,$q$}}
\psfrag{"inbp2.dat"}[Cr][Cr]{\tiny{$\tilde{p}$,$\tilde{q}$}}
\psfrag{"inbetaL.dat"}[Cr][Cr]{\tiny{$\ve$,$\tilde{\ve}$}}
\psfrag{"inbbetaL.dat"}[Cr][Cr]{\tiny{$\ve'$,$\tilde{\ve}'$}}
\psfrag{"inbE1.dat"}[Cr][Cr]{\tiny{$E$}}
\psfrag{"inbE2.dat"}[Cr][Cr]{\tiny{$\tilde{E}$}}
\includegraphics*[width=1.0\columnwidth]{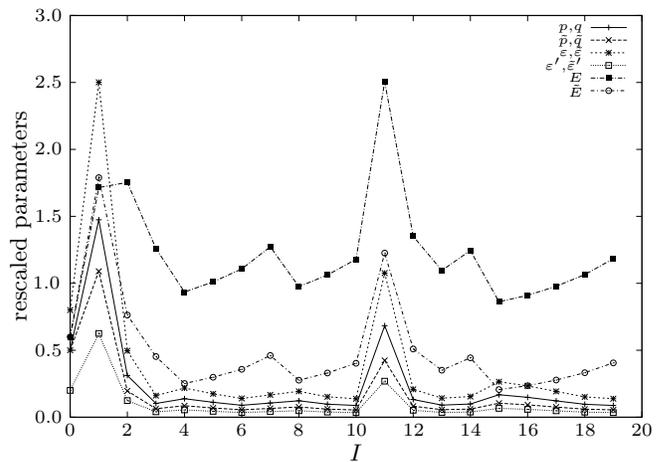}
\caption{Evolution of parameters in Eqs.~\eqref{e:alpha2} and~\eqref{e:gamma2} under $b=2$ decimation.  Pure case with $\omega=0.41$ and starting parameter values $p=\tilde{p}=q=\tilde{q}=0.5$, $\ve=\tilde{\ve}=0.8$, $\ve'=\tilde{\ve}'=0.2$, $E=\tilde{E}=0.6$.  Graph shows absolute magnitude of parameters against number of iterations $I$; lines are provided as an aid to the eye. Chaotic behaviour is typical of scaling in the allowed band.}
\label{f:inband}
\end{center}
\end{figure}

\begin{figure}
\begin{center}
\psfrag{I}[Tc][Bc]{$I$}
\psfrag{r}[Bc][Tc]{$\text{rescaled parameters}$}
\psfrag{0}[Cr][Cr]{\scriptsize{0.0}}
\psfrag{0.5}[Cr][Cr]{\scriptsize{0.5}}
\psfrag{1}[Cr][Cr]{\scriptsize{1.0}}
\psfrag{1.5}[Cr][Cr]{\scriptsize{1.5}}
\psfrag{2}[Cr][Cr]{\scriptsize{2.0}}
\psfrag{2.5}[Cr][Cr]{\scriptsize{2.5}}
\psfrag{3}[Cr][Cr]{\scriptsize{3.0}}
\psfrag{0b}[Tc][Tc]{\scriptsize{0}}
\psfrag{2b}[Tc][Tc]{\scriptsize{2}}
\psfrag{4}[Tc][Tc]{\scriptsize{4}}
\psfrag{6}[Tc][Tc]{\scriptsize{6}}
\psfrag{8}[Tc][Tc]{\scriptsize{8}}
\psfrag{10}[Tc][Tc]{\scriptsize{10}}
\psfrag{12}[Tc][Tc]{\scriptsize{12}}
\psfrag{14}[Tc][Tc]{\scriptsize{14}}
\psfrag{16}[Tc][Tc]{\scriptsize{16}}
\psfrag{18}[Tc][Tc]{\scriptsize{18}}
\psfrag{20}[Tc][Tc]{\scriptsize{20}}
\psfrag{"outbp1.dat"}[Cr][Cr]{\tiny{$p$,$q$}}
\psfrag{"outbp2.dat"}[Cr][Cr]{\tiny{$\tilde{p}$,$\tilde{q}$}}
\psfrag{"outbetaL.dat"}[Cr][Cr]{\tiny{$\ve$,$\tilde{\ve}$}}
\psfrag{"outbbetaL.dat"}[Cr][Cr]{\tiny{$\ve'$,$\tilde{\ve}'$}}
\psfrag{"outbE1.dat"}[Cr][Cr]{\tiny{$E$}}
\psfrag{"outbE2.dat"}[Cr][Cr]{\tiny{$\tilde{E}$}}
\includegraphics*[width=1.0\columnwidth]{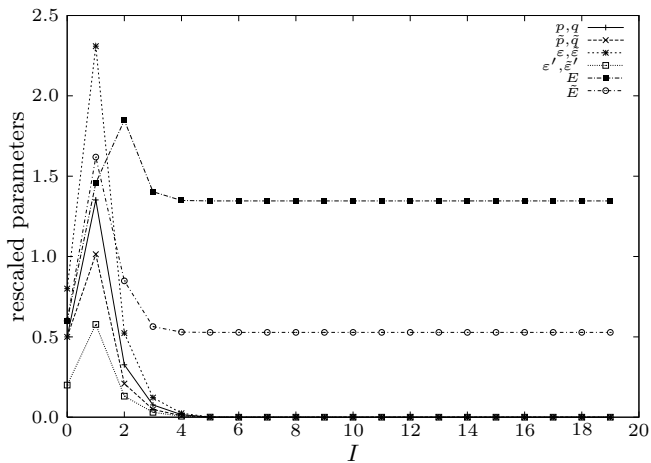}
\caption{Evolution of parameters in Eqs.~\eqref{e:alpha2} and~\eqref{e:gamma2} under $b=2$ decimation.  Pure case with $\omega=0.39$ and starting parameter values $p=\tilde{p}=q=\tilde{q}=0.5$, $\ve=\tilde{\ve}=0.8$, $\ve'=\tilde{\ve}'=0.2$, $E=\tilde{E}=0.6$.  Graph shows absolute magnitude of parameters against number of iterations $I$; lines are provided as an aid to the eye. Rate parameters decay exponentially as a function of $2^I$, as expected outside the allowed band.}
\label{f:outband}
\end{center}
\end{figure}

For $2\ve' \leq \omega \leq 2\ve$ the scaling parameters all evolve chaotically under scaling (see Fig.~\ref{f:inband}).
In contrast for $\omega <  2\ve'$ or $\omega > 2\ve$, the ``energies'' $E$ and $\tilde{E}$ tend to constant values whereas the rescaled rates $r$ (i.e, $p$, $q$, etc.) tend rapidly to zero inside an exponentially decaying envelope (see Fig.~\ref{f:outband}), i.e., $r(l) \sim f(l) e^{-l/\xi}$ where  $l=2^I$ is the distance between sites ($I$ is the number of iterations) and  $\xi$ is a localization length. 
We also see this localized behaviour for all non-real $\omega$.

\begin{figure}
\begin{center}
\psfrag{w}[Tc][Bc]{$\omega$}
\psfrag{l}[Bc][Tc]{$\xi$}
\psfrag{0b}[Tc][Tc]{\scriptsize{0.0}}
\psfrag{0.5}[Tc][Tc]{\scriptsize{0.5}}
\psfrag{1b}[Tc][Tc]{\scriptsize{1.0}}
\psfrag{1.5}[Tc][Tc]{\scriptsize{1.5}}
\psfrag{2b}[Tc][Tc]{\scriptsize{2.0}}
\psfrag{0}[Cr][Cr]{\scriptsize{0}}
\psfrag{1}[Cr][Cr]{\scriptsize{1}}
\psfrag{2}[Cr][Cr]{\scriptsize{2}}
\psfrag{3}[Cr][Cr]{\scriptsize{3}}
\psfrag{4}[Cr][Cr]{\scriptsize{4}}
\psfrag{5}[Cr][Cr]{\scriptsize{5}}
\psfrag{"p0805finp1locre.dat"}[Cr][Cr]{\tiny{numerical scaling value of $\xi$}}
\psfrag{f(x)}[Cr][Cr]{\tiny{analytical prediction for $\xi$}}
\includegraphics*[width=1.0\columnwidth]{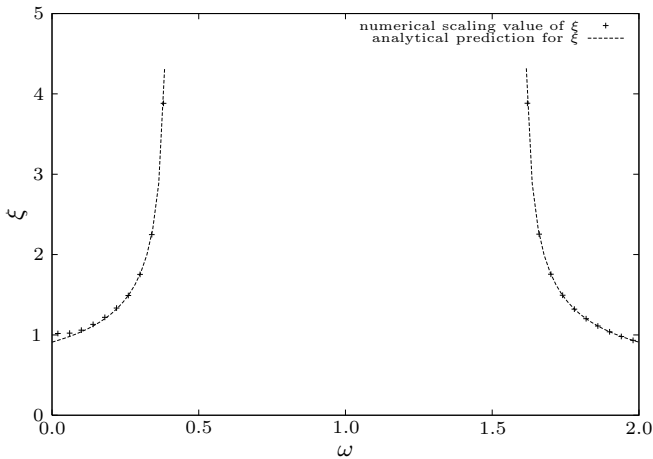}
\caption{Localization length $\xi$ versus (real) frequency $\omega$ for pure case with $p=q=0.5$, $\ve=0.8$, $\ve'=0.2$.  $\xi$ is effectively infinite in allowed band $0.4<\omega<1.6$.  Crosses indicate data from numerical scaling of rate $p$; dashed line is analytical prediction of Eq.~\eqref{e:loclsym}. 
}
\label{f:ploclength}
\end{center}
\end{figure}

Just as in the scaling of the Cole-Hopf-transformed mean-field ASEP~\cite{Me04} the localization length can be measured numerically from the exponential decay of the parameters.  In Fig.~\ref{f:ploclength} we plot the localization length obtained from the scaling of $p$, as a function of real frequency $\omega$ (localization lengths obtained from the other rates are found to be identical).
These numerical results can be simply explained by reference to the pure spectrum for the symmetric case.
The ``allowed'' band (real $k$) corresponds to the region $2\ve' \leq \omega \leq 2\ve$; here states are extended and the localization length is infinite.  Frequencies outside this band correspond to a complex wave-vector $k+i\kappa$, i.e., localized states with localization length $\xi = 1/\kappa$.  Indeed, from $\omega = \ve + \ve' + (\ve - \ve') \cos (k+i\kappa)$ we can obtain an analytical expression for the localization length on the real-$\omega$ axis ($k=0$ or $k=\pi$):
\begin{equation}
\xi=\left( \text{arccosh} \left| \frac{\ve'+\ve-\omega}{\ve'-\ve} \right| \right)^{-1} \label{e:loclsym}.
\end{equation}
This expression is shown as a dashed line in Fig.~\ref{f:ploclength} and agreement with the numerics is excellent.

\subsubsection{Biased case, $p \neq q$}

\begin{figure}
\begin{center}
\psfrag{(p-q)}[Cr][Cr]{$(p-q)$}
\psfrag{-(p-q)}[Cr][Cr]{$-(p-q)$}
\psfrag{Y}[Br][Br]{$\text{Im} \lambda_k$}
\psfrag{2e}[Br][Br]{$2\varepsilon'$}
\psfrag{2f}[Bl][Bl]{$2\varepsilon$}
\psfrag{1}[Bl][Bl]{$\ve+\ve'$}
\psfrag{X}[Tl][Tl]{$\text{Re} \lambda_k$}
\psfrag{Ballistic}[][]{ballistic}
\psfrag{Localized}[][]{localized}
\includegraphics*[width=0.8\columnwidth]{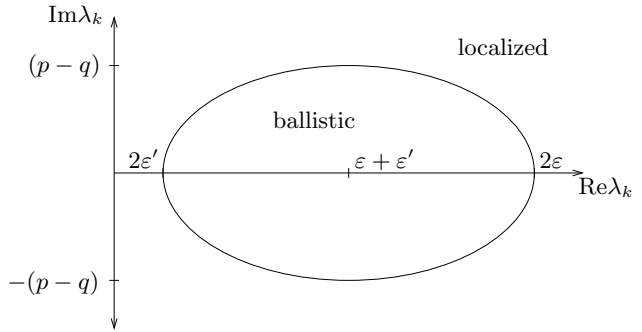}
\caption{$\text{Im} \lambda_k$ versus $\text{Re} \lambda_k$ for pure model (case $p>q$, $\ve>\ve'$).  Solid ellipse represents allowed spectrum given by~\eqref{e:spectrum}.  Inside the ellipse equations of motion scale to ballistic limit $p/q \to \infty$; outside of ellipse is localized region.}
\label{f:purespec}
\end{center}
\end{figure}

In the pure biased case the ``allowed'' values of $\omega$ are given by $\lambda_k$ of Eq.~\eqref{e:spectrum} with $k$ real---an ellipse in complex $\lambda_k$-space, see Fig.~\ref{f:purespec}.  
For these values of $\omega$ we again find, as expected, that the scaling parameters all evolve chaotically under scaling corresponding to extended states.  Outside this allowed ellipse we see localized states once more.  

\begin{figure}
\begin{center}
\psfrag{I}[Tc][Bc]{$I$}
\psfrag{r}[Bc][Tc]{$\text{rescaled parameters}$}
\psfrag{0b}[Tr][Tr]{\scriptsize{0}}
\psfrag{1}[Tc][Tc]{\scriptsize{1}}
\psfrag{2}[Tc][Tc]{\scriptsize{2}}
\psfrag{3}[Tc][Tc]{\scriptsize{3}}
\psfrag{4}[Tc][Tc]{\scriptsize{4}}
\psfrag{5b}[Tc][Tc]{\scriptsize{5}}
\psfrag{0}[Tc][Tc]{\scriptsize{0}}
\psfrag{5}[Cr][Cr]{\scriptsize{5}}
\psfrag{10}[Cr][Cr]{\scriptsize{10}}
\psfrag{15}[Cr][Cr]{\scriptsize{15}}
\psfrag{20}[Cr][Cr]{\scriptsize{20}}
\psfrag{25}[Cr][Cr]{\scriptsize{25}}
\psfrag{30}[Cr][Cr]{\scriptsize{30}}
\psfrag{35}[Cr][Cr]{\scriptsize{35}}
\psfrag{40}[Cr][Cr]{\scriptsize{40}}
\psfrag{45}[Cr][Cr]{\scriptsize{45}}
\psfrag{50}[Cr][Cr]{\scriptsize{50}}
\psfrag{"ep1.dat"}[Cl][Cl]{\tiny{$p$}}
\psfrag{"ep2.dat"}[Cl][Cl]{\tiny{$\tilde{p}$}}
\psfrag{"eetaL.dat"}[Cl][Cl]{\tiny{$\ve$}}
\psfrag{"ebetaL.dat"}[Cl][Cl]{\tiny{$\ve'$}}
\psfrag{"eq1.dat"}[Cl][Cl]{\tiny{$q$}}
\psfrag{"eq2.dat"}[Cl][Cl]{\tiny{$\tilde{q}$}}
\psfrag{"eetaR.dat"}[Cl][Cl]{\tiny{$\tilde{\ve}$}}
\psfrag{"ebetaR.dat"}[Cl][Cl]{\tiny{$\tilde{\ve}'$}}
\psfrag{"eE1.dat"}[Cl][Cl]{\tiny{$E$}}
\psfrag{"eE2.dat"}[Cl][Cl]{\tiny{$\tilde{E}$}}
\includegraphics*[width=1.0\columnwidth]{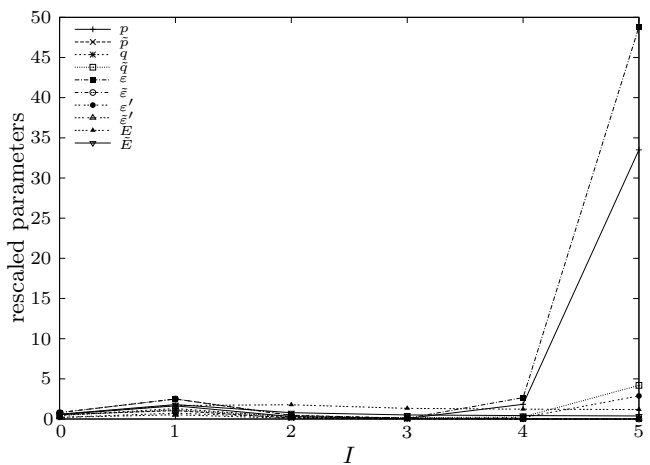}
\caption{Evolution of parameters in Eqs.~\eqref{e:alpha2} and~\eqref{e:gamma2} under $b=2$ decimation (first few iterations).  Pure case with $\omega=0.41$ and starting parameter values $p=\tilde{p}=0.55$, $q=\tilde{q}=0.45$, $\ve=\tilde{\ve}=0.8$, $\ve'=\tilde{\ve}'=0.2$, $E=\tilde{E}=0.6$.  Graph shows absolute magnitude of parameters against number of iterations $I$; lines are provided as an aid to the eye.  Rates $p$, $\tilde{q}$, $\ve$, $\ve'$ scale exponentially to infinity while $\tilde{p}$, $q$, $\tilde{\ve}$, $\tilde{\ve}'$ decay exponentially and ``energies'' $E$, $\tilde{E}$ tend to constant values.  This is typical of ballistic behaviour inside the allowed ellipse (for case $p>q$).  
}
\label{f:inellipse}
\end{center}
\end{figure}

Inside the ellipse we see a new type of behaviour---$E$ and $\tilde{E}$ again tend to constant values, some of the rates scale exponentially to zero, whereas others grow exponentially to infinity (see Fig.~\ref{f:inellipse}).  
We argue that in this region the system is dominated by the ballistic motion and scales to the totally asymmetric case.  So if we start with $p>q$, then $p$ scales to infinity while $q$ decays to zero.  

\begin{figure}
\begin{center}
\psfrag{w}[Tc][Bc]{$\omega_{\text{Re}}$}
\psfrag{l}[Bc][Tc]{$\xi$}
\psfrag{0b}[Tc][Tc]{\scriptsize{0.0}}
\psfrag{0.5}[Tc][Tc]{\scriptsize{0.5}}
\psfrag{1b}[Tc][Tc]{\scriptsize{1.0}}
\psfrag{1.5}[Tc][Tc]{\scriptsize{1.5}}
\psfrag{2b}[Tc][Tc]{\scriptsize{2.0}}
\psfrag{0}[Cr][Cr]{\scriptsize{0}}
\psfrag{5}[Cr][Cr]{\scriptsize{5}}
\psfrag{10}[Cr][Cr]{\scriptsize{10}}
\psfrag{15}[Cr][Cr]{\scriptsize{15}}
\psfrag{20}[Cr][Cr]{\scriptsize{20}}
\psfrag{25}[Cr][Cr]{\scriptsize{25}}
\psfrag{-5}[Cr][Cr]{\scriptsize{-5}}
\psfrag{-10}[Cr][Cr]{\scriptsize{-10}}
\psfrag{-15}[Cr][Cr]{\scriptsize{-15}}
\psfrag{-20}[Cr][Cr]{\scriptsize{-20}}
\psfrag{-25}[Cr][Cr]{\scriptsize{-25}}
\psfrag{"p08055refinp1locre.dat"}[Cr][Cr]{\tiny{$\xi_p$}}
\psfrag{"p08055refinq1locre.dat"}[Cr][Cr]{\tiny{$\xi_q$}}
\includegraphics*[width=1.0\columnwidth]{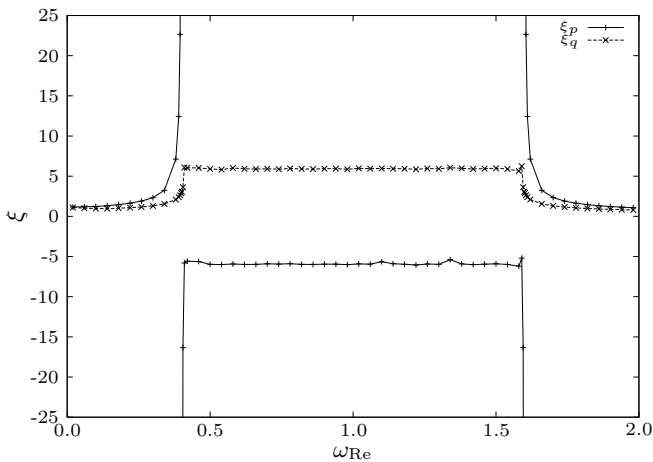}
\caption{Numerical localization lengths $\xi_p$, $\xi_q$ as a function of real frequency $\omega_{\text{Re}}$ for pure case with $p=0.55$,  $q=0.45$, $\ve=0.8$, $\ve'=0.2$.   Note difference in scaling behaviour for $p$ and $q$.   Lines are provided as an aid to the eye.} 
\label{f:locre}
\end{center}
\end{figure}

\begin{figure}
\begin{center}
\psfrag{w}[Tc][Bc]{$\omega_{\text{Im}}$}
\psfrag{l}[Bc][Tc]{$\xi$}
\psfrag{0b}[Tc][Tc]{\scriptsize{0.0}}
\psfrag{0.1}[Tc][Tc]{\scriptsize{0.1}}
\psfrag{0.2}[Tc][Tc]{\scriptsize{0.2}}
\psfrag{0.3}[Tc][Tc]{\scriptsize{0.3}}
\psfrag{0.4}[Tc][Tc]{\scriptsize{0.4}}
\psfrag{-0.1}[Tc][Tc]{\scriptsize{-0.1}}
\psfrag{-0.2}[Tc][Tc]{\scriptsize{-0.2}}
\psfrag{-0.3}[Tc][Tc]{\scriptsize{-0.3}}
\psfrag{-0.4}[Tc][Tc]{\scriptsize{-0.4}}
\psfrag{0}[Cr][Cr]{\scriptsize{0}}
\psfrag{50}[Cr][Cr]{\scriptsize{50}}
\psfrag{100}[Cr][Cr]{\scriptsize{100}}
\psfrag{150}[Cr][Cr]{\scriptsize{150}}
\psfrag{-50}[Cr][Cr]{\scriptsize{-50}}
\psfrag{-100}[Cr][Cr]{\scriptsize{-100}}
\psfrag{-150}[Cr][Cr]{\scriptsize{-150}}
\psfrag{"p08055imfinp1locim.dat"}[Cr][Cr]{\tiny{$\xi_p$}}
\psfrag{"p08055imfinq1locim.dat"}[Cr][Cr]{\tiny{$\xi_q$}}
\includegraphics*[width=1.0\columnwidth]{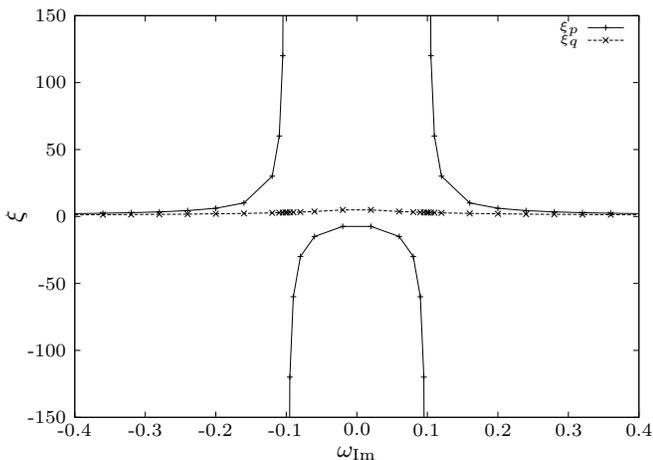}
\caption{Numerical localization lengths $\xi_p$, $\xi_q$ as a function of complex frequency $\omega = 1 + i \omega_{\text{Im}}$ for pure case with $p=0.55$,  $q=0.45$, $\ve=0.8$, $\ve'=0.2$.   Note difference in scaling behaviour for $p$ and $q$.   Lines are provided as an aid to the eye.} 
\label{f:locim}
\end{center}
\end{figure}

In Figs.~\ref{f:locre} and~\ref{f:locim} we plot the localization lengths for $p$ and $q$ along the principal axes of the $\omega$-space ellipse (again for the case $p>q$).  
As is evident from the discussion above, the localization length for $q$ is small and positive in all regions---$q$ is an irrelevant parameter under scaling.  The localization length for $p$ is positive in the localized region outside the ellipse, infinite on the allowed ellipse, and negative in the ballistic region inside (refer again to Fig.~\ref{f:purespec}).  The value of this relevant localization length can again be understood analytically by considering a complex wavevector in the dispersion relation.

\subsection{Disordered model}
\label{ss:ffdis}

Armed with this understanding of the pure case, we move on to look at the effect of disorder on the free-fermion system.
There are obviously various different possible ways to add disorder; here we 
mainly investigate the case where the disorder is in $\ve$ and $\ve'$ with the rates chosen to maintain the free-fermion condition for each bond (specifically $\ve'_j=1-\ve_j$, $p_j=p$ and $q_j=q$, where $p+q=1$).
Once again, we consider in turn the symmetric ($p=q$) and asymmetric ($p \neq q$) cases.

\subsubsection{Non-biased case, $p=q$}

\begin{figure}
\begin{center}
\psfrag{I}[Tc][Bc]{$I$}
\psfrag{r}[Bc][Tc]{$\text{rescaled parameters}$}
\psfrag{0b}[Tr][Tr]{\scriptsize{0}}
\psfrag{2b}[Tc][Tc]{\scriptsize{2}}
\psfrag{4b}[Tc][Tc]{\scriptsize{4}}
\psfrag{6b}[Tc][Tc]{\scriptsize{6}}
\psfrag{8b}[Tc][Tc]{\scriptsize{8}}
\psfrag{10b}[Tc][Tc]{\scriptsize{10}}
\psfrag{12b}[Tc][Tc]{\scriptsize{12}}
\psfrag{0}[Cr][Cr]{\scriptsize{0}}
\psfrag{2}[Cr][Cr]{\scriptsize{2}}
\psfrag{4}[Cr][Cr]{\scriptsize{4}}
\psfrag{6}[Cr][Cr]{\scriptsize{6}}
\psfrag{8}[Cr][Cr]{\scriptsize{8}}
\psfrag{10}[Cr][Cr]{\scriptsize{10}}
\psfrag{12}[Cr][Cr]{\scriptsize{12}}
\psfrag{14}[Cr][Cr]{\scriptsize{14}}
\psfrag{16}[Cr][Cr]{\scriptsize{16}}
\psfrag{"disp1.dat"}[Cr][Cr]{\tiny{$p$}} 
\psfrag{"disq1.dat"}[Cr][Cr]{\tiny{$q$}} 
\psfrag{"disp2.dat"}[Cr][Cr]{\tiny{$\tilde{p}$}} 
\psfrag{"disq2.dat"}[Cr][Cr]{\tiny{$\tilde{q}$}} 
\psfrag{"disetaL.dat"}[Cr][Cr]{\tiny{$\ve$}} 
\psfrag{"disetaR.dat"}[Cr][Cr]{\tiny{$\tilde{\ve}$}} 
\psfrag{"disbetaL.dat"}[Cr][Cr]{\tiny{$\ve'$}} 
\psfrag{"disbetaR.dat"}[Cr][Cr]{\tiny{$\tilde{\ve}'$}} 
\psfrag{"disE1.dat"}[Cr][Cr]{\tiny{$E$}} 
\psfrag{"disE2.dat"}[Cr][Cr]{\tiny{$\tilde{E}$}} 
\includegraphics*[width=1.0\columnwidth]{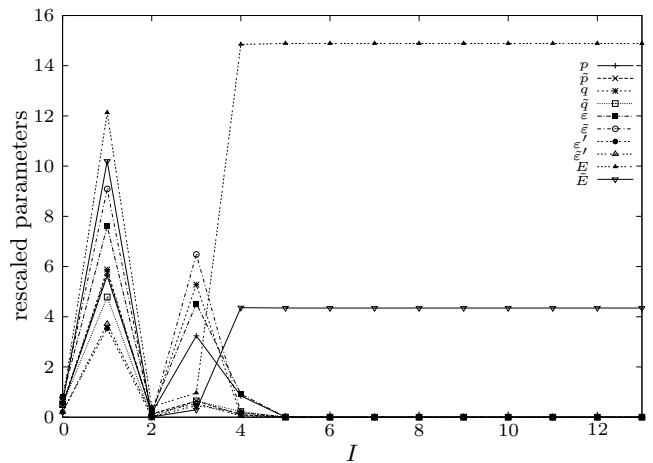}
\caption{Evolution of parameters in Eqs.~\eqref{e:alpha2}and~\eqref{e:gamma2} under $b=2$ decimation.  Disordered case with $\omega=1.41$ and starting parameter values $p_{j,j-1}=\tilde{p}_{j-1,j}=q_{j-1,j}=\tilde{q}_{j,j-1}=0.5$, $\ve_{j,j-1}=\tilde{\ve}_{j-1,j}=z$, $\ve'_{j,j-1}=\tilde{\ve}'_{j-1,j}=1-z$, $E=\tilde{E}=p_{j,j-1}-p_{j+1,j}-\ve_{j+1,j}-\ve'_{j,j-1}$ where $z$ is a random variable drawn from a uniform distribution between 0.6 and 1.0.  Graph shows absolute magnitude of parameters at site $j_0$ (averaged over 100 realizations of disorder) against number of iterations $I$; lines are provided as an aid to the eye. Rate parameters decay exponentially as a function of $2^I$ showing a clear localization effect.}
\label{f:indis}
\end{center}
\end{figure}

\begin{figure}
\begin{center}
\psfrag{w}[Tc][Bc]{$\omega$}
\psfrag{l}[Bc][Tc]{$\xi$}
\psfrag{0b}[Tc][Tc]{\scriptsize{0.0}}
\psfrag{0.5}[Tc][Tc]{\scriptsize{0.5}}
\psfrag{1}[Tc][Tc]{\scriptsize{1.0}}
\psfrag{1.5}[Tc][Tc]{\scriptsize{1.5}}
\psfrag{2}[Tc][Tc]{\scriptsize{2.0}}
\psfrag{0.2}[Tc][Tc]{\scriptsize{0.2}}
\psfrag{0.4}[Tc][Tc]{\scriptsize{0.4}}
\psfrag{0.6}[Tc][Tc]{\scriptsize{0.6}}
\psfrag{0.8}[Tc][Tc]{\scriptsize{0.8}}
\psfrag{1.2}[Tc][Tc]{\scriptsize{1.2}}
\psfrag{1.4}[Tc][Tc]{\scriptsize{1.4}}
\psfrag{1.6}[Tc][Tc]{\scriptsize{1.6}}
\psfrag{1.8}[Tc][Tc]{\scriptsize{1.8}}
\psfrag{0}[Cr][Cr]{\scriptsize{0}}
\psfrag{50}[Cr][Cr]{\scriptsize{50}}
\psfrag{100}[Cr][Cr]{\scriptsize{100}}
\psfrag{150}[Cr][Cr]{\scriptsize{150}}
\psfrag{200}[Cr][Cr]{\scriptsize{200}}
\psfrag{250}[Cr][Cr]{\scriptsize{250}}
\psfrag{"p0805finp1locre.dat"}[Cr][Cr]{\tiny{pure case}}
\psfrag{"D0805finp1locre.dat"}[Cr][Cr]{\tiny{disordered case}}
\includegraphics*[width=1.0\columnwidth]{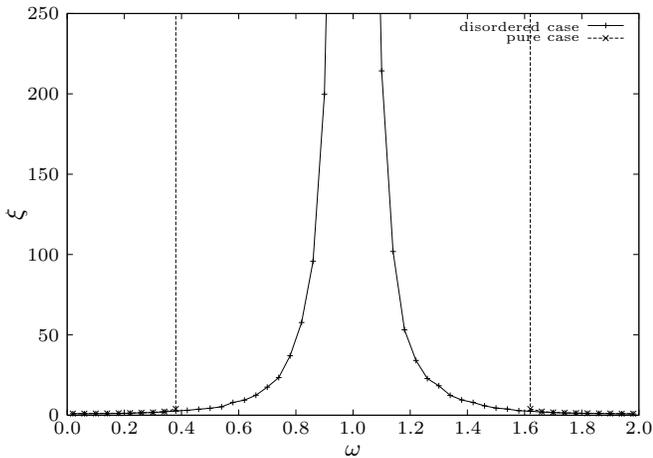}
\caption{Localization length for $p$ scaling versus (real) frequency for disordered case with $p=q=0.5$, $\ve_j=z$, $\ve'_j=1-z$ where $z$ is drawn from a uniform distribution between 0.6 and 1.0.  Pure results outside the band are also shown for comparison (dashed). In the disordered case, the localization length is finite everywhere except $\omega=1$.} 
\label{f:disloc}
\end{center}
\end{figure}

In Fig.~\ref{f:indis} we show an example of the scaling behaviour for the case with $p_j=q_j=0.5$, $\ve_j$ drawn from a uniform distribution and $\ve'_j=1-\ve_j$.  The value of $\omega$ corresponds to being in the allowed band for the pure case (compare Fig.~\ref{f:inband}). 
We find that the rates evolve chaotically towards zero within an exponential envelope and so we can define a localization length via $r(l) \sim f(l) e^{-l/\xi}$.  In Fig.~\ref{f:disloc} we plot the localization length for the decay of $p$, averaged over realizations of disorder.   
Localization is clearly seen for all values of $\omega$ in the pure band.

\subsubsection{Biased case, $p \neq q$}

\begin{figure}
\begin{center}
\psfrag{w}[Tc][Bc]{$\omega_{\text{Re}}$}
\psfrag{l}[Bc][Tc]{$\xi$}
\psfrag{0b}[Tc][Tc]{\scriptsize{0.0}}
\psfrag{0.5}[Tc][Tc]{\scriptsize{0.5}}
\psfrag{1}[Tc][Tc]{\scriptsize{1.0}}
\psfrag{1.5}[Tc][Tc]{\scriptsize{1.5}}
\psfrag{2}[Tc][Tc]{\scriptsize{2.0}}
\psfrag{0.2}[Tc][Tc]{\scriptsize{0.2}}
\psfrag{0.4}[Tc][Tc]{\scriptsize{0.4}}
\psfrag{0.6}[Tc][Tc]{\scriptsize{0.6}}
\psfrag{0.8}[Tc][Tc]{\scriptsize{0.8}}
\psfrag{1.2}[Tc][Tc]{\scriptsize{1.2}}
\psfrag{1.4}[Tc][Tc]{\scriptsize{1.4}}
\psfrag{1.6}[Tc][Tc]{\scriptsize{1.6}}
\psfrag{1.8}[Tc][Tc]{\scriptsize{1.8}}
\psfrag{0}[Cr][Cr]{\scriptsize{0}}
\psfrag{50}[Cr][Cr]{\scriptsize{50}}
\psfrag{100}[Cr][Cr]{\scriptsize{100}}
\psfrag{-50}[Cr][Cr]{\scriptsize{-50}}
\psfrag{-100}[Cr][Cr]{\scriptsize{-100}}
\psfrag{"Drep1locre.dat"}[Cr][Cr]{\tiny{disordered, $\xi_p$}}
\psfrag{"Dreq1locre.dat"}[Cr][Cr]{\tiny{disordered, $\xi_q$}}
\psfrag{"p08055refinp1locre.dat"}[Cr][Cr]{\tiny{pure, $\xi_p$}}
\psfrag{"p08055refinq1locre.dat"}[Cr][Cr]{\tiny{pure, $\xi_q$}}
\includegraphics*[width=1.0\columnwidth]{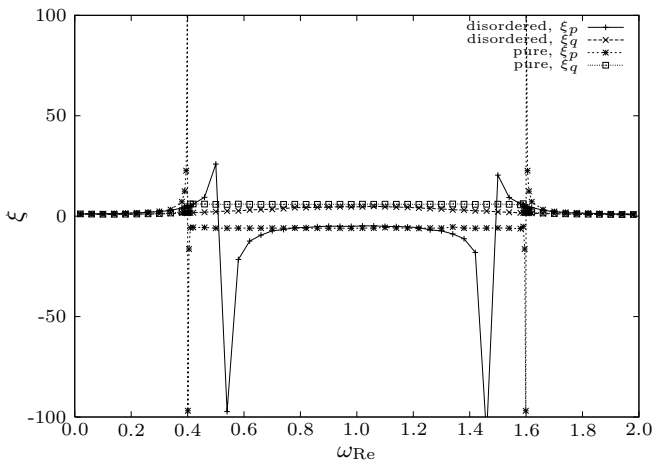}
\caption{Numerical localization lengths $\xi_p$, $\xi_q$ as a function of real frequency $\omega_{\text{Re}}$ for disordered case with $p=0.55$, $q=0.45$, $\ve_j=z$, $\ve'_j=1-z$ where $z$ is drawn from a uniform distribution between 0.6 and 1.0.  Lines are provided as an aid to the eye. Pure results are shown for comparison.  Note the reduction of the ballistic region in the disordered case.} 
\label{f:locdre}
\end{center}
\end{figure}

In Fig.~\ref{f:locdre} we plot the localization lengths for $p$ and $q$ along the real frequency axis for a sample disordered case, viz., $p_j=0.55$, $q_j=0.45$, $\ve_j$ drawn from a uniform distribution and $\ve'_j=1-\ve_j$.  The pure results are also shown for reference (compare also Fig.~\ref{f:locre}).
In this case we see that disorder reduces the size of the allowed ellipse along the real-$\omega$ axis, i.e., it acts to increase the localized region at the expense of the ballistic region.  The results for $\omega = 1 + i \omega_{\text{Im}}$ are essentially the same as in the pure case (Fig.~\ref{f:locim}).

This competition between disorder-induced localization and driving is analogous to that arising in another non-Hermitian Hamiltonian studied by Hatano and Nelson~\cite{Hatano96}.  Their work was motivated by the mapping between flux lines in a ($d$+1)-dimensional superconductor and $d$-dimensional bosons.   A random potential in the bosonic problem arises from the study of columnar defects in the superconductor.

\begin{figure}
\begin{center}
\psfrag{w}[Tc][Bc]{$\omega$}
\psfrag{l}[Bc][Tc]{$\xi_p$}
\psfrag{0.36}[Tc][Tc]{\scriptsize{0.36}}
\psfrag{0.38}[Tc][Tc]{\scriptsize{0.38}}
\psfrag{0.4}[Tc][Tc]{\scriptsize{0.40}}
\psfrag{0.42}[Tc][Tc]{\scriptsize{0.42}}
\psfrag{0.44}[Tc][Tc]{\scriptsize{0.44}}
\psfrag{0.46}[Tc][Tc]{\scriptsize{0.46}}
\psfrag{0.48}[Tc][Tc]{\scriptsize{0.48}}
\psfrag{0.5}[Tc][Tc]{\scriptsize{0.50}}
\psfrag{0}[Cr][Cr]{\scriptsize{0}}
\psfrag{100}[Cr][Cr]{\scriptsize{100}}
\psfrag{200}[Cr][Cr]{\scriptsize{200}}
\psfrag{300}[Cr][Cr]{\scriptsize{300}}
\psfrag{400}[Cr][Cr]{\scriptsize{400}}
\psfrag{500}[Cr][Cr]{\scriptsize{500}}
\psfrag{600}[Cr][Cr]{\scriptsize{600}}
\psfrag{700}[Cr][Cr]{\scriptsize{700}}
\psfrag{800}[Cr][Cr]{\scriptsize{800}}
\psfrag{900}[Cr][Cr]{\scriptsize{900}}
\psfrag{1000}[Cr][Cr]{\scriptsize{1000}}
\psfrag{-100}[Cr][Cr]{\scriptsize{-100}}
\psfrag{-200}[Cr][Cr]{\scriptsize{-200}}
\psfrag{-300}[Cr][Cr]{\scriptsize{-300}}
\psfrag{-400}[Cr][Cr]{\scriptsize{-400}}
\psfrag{-500}[Cr][Cr]{\scriptsize{-500}}
\psfrag{-600}[Cr][Cr]{\scriptsize{-600}}
\psfrag{-700}[Cr][Cr]{\scriptsize{-700}}
\psfrag{-800}[Cr][Cr]{\scriptsize{-800}}
\psfrag{-900}[Cr][Cr]{\scriptsize{-900}}
\psfrag{-1000}[Cr][Cr]{\scriptsize{-1000}}
\psfrag{"nnD0608p1locre.dat"}[Cr][Cr]{\tiny{$p=0.6$, $q=0.4$}}
\psfrag{"nnD0708p1locre.dat"}[Cr][Cr]{\tiny{$p=0.7$, $q=0.3$}}
\psfrag{"nnD0808p1locre.dat"}[Cr][Cr]{\tiny{$p=0.8$, $q=0.2$}}
\psfrag{"nnD0908p1locre.dat"}[Cr][Cr]{\tiny{$p=0.9$, $q=0.1$}}
\psfrag{"nnD1008p1locre.dat"}[Cr][Cr]{\tiny{$p=1.0$, $q=0.0$}}
\includegraphics*[width=1.0\columnwidth]{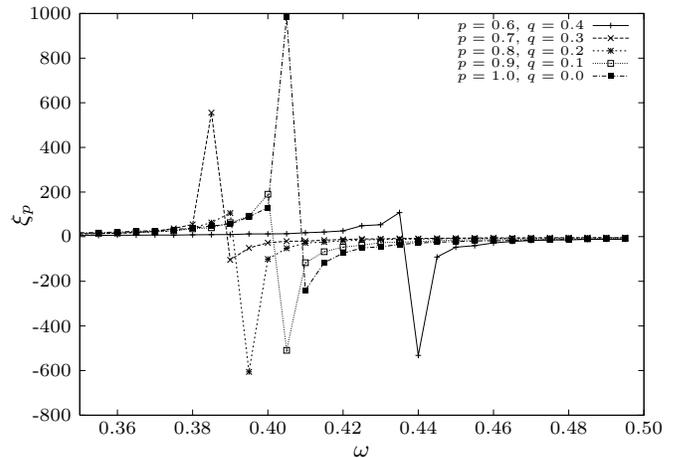}
\caption{Numerical localization length $\xi_p$ as a function of real frequency $\omega_\text{Re}$ for disordered cases with $\ve_j=z$, $\ve'_j=1-z$ where $z$ is drawn from a uniform distribution between 0.6 and 1.0.  Results are shown for a range of values of $p$ and $q$ as indicated by the legend.  Lines are provided as an aid to the eye.} 
\label{f:pqrange}
\end{center}
\end{figure}

For all cases with disorder only in $\ve$ and $\ve'$, we find that the localization length is altered along the real-$\omega$ axis but unchanged along the axis in the imaginary direction.\footnote{Obviously disorder in $p$ and $q$ affects the imaginary-$\omega$ direction instead.}  However, the change in the position of the allowed band (where the localization length is infinite) depends on the strength of the driving $p-q$, see Fig.~\ref{f:pqrange}.
We shall now consider some tentative analytical ideas to quantify these disorder effects.

\begin{figure}
\begin{center}
\psfrag{(p-q)}[Cr][Cr]{$(p-q)$}
\psfrag{-(p-q)}[Cr][Cr]{$-(p-q)$}
\psfrag{Y}[Br][Br]{$\text{Im} \lambda_k$}
\psfrag{2e}[Bl][Tr]{\tiny{$\frac{b_0+a_0}{2}$}}
\psfrag{2f}[Br][Tl]{\tiny{$\frac{b_0-a_0}{2}$}}
\psfrag{1}[Bl][Bl]{$b_0$}
\psfrag{X}[Tl][Tl]{$\text{Re} \lambda_k$}
\psfrag{Ballistic}[][]{ballistic}
\psfrag{Localized}[][]{localized}
\includegraphics*[width=0.8\columnwidth]{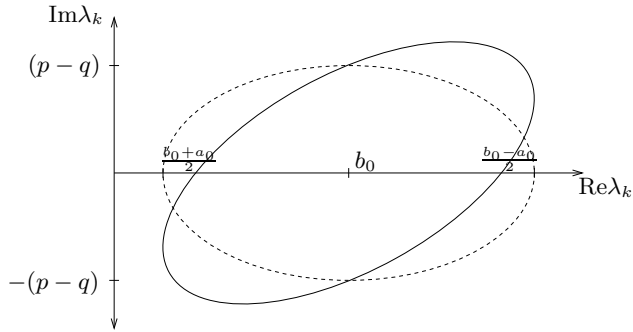}
\caption{Schematic of effective disordered dispersion relation in $\omega$-space for $\lambda_k = b_0 + (a_0 + i 2 \delta) \cos k + i(p-q)\sin k$ (solid line).  Pure case $\lambda_k = b_0 + a_0 \cos k + i(p-q)\sin k$ is shown for comparison (dashed line).  Inside the ellipse equations of motion scale to ballistic limit $p/q \to \infty$; outside of ellipse is localized region.}
\label{f:disspec}
\end{center}
\end{figure}

Recall that the pure spectrum is given by
\begin{equation}
\lambda_k=b_0 - a_0 \cos k + i(p-q)\sin k,
\end{equation}
with $b_0=\ve'+\ve$ and $a_0=\ve'-\ve$.  We seek to modify this to represent the effective spectrum of the disordered model.  For the case of disordered $\ve$ and $\ve'$ (with $\ve_j+\ve'_j=p+q=1$) then a naive suggestion is
\begin{equation}
\lambda_k=b_0 - a \cos k + i(p-q)\sin k \label{e:disspec}
\end{equation}
where $a$ can, in principle, have both real and imaginary parts, i.e.,
\begin{equation}
a=\tilde{a}+i2\delta
\end{equation}
with $\tilde{a}$, $\delta$ real.  The imaginary part of $a$ causes a rotation of the allowed ellipse (see Fig.~\ref{f:disspec}); if $\tilde{a} \neq a_0$, then there is also stretching or compression.  
In agreement with the numerical localization length data, this form leaves the centre of the ellipse and the intersection with the $\omega=b_0$ axis unchanged,   It also explains why any amount of disorder causes localization in the $p=q$ case (where the ellipse is a line along the real axis) for all values of real frequency except $\omega=1$. 

There is no \emph{a priori} reason why the imaginary part of $a$ should have a positive sign so we naturally expect the possibility of both $\pm 2\delta$.  Furthermore, we anticipate that $\tilde{a}$ is related to the mean of $\ve'-\ve$ and $\delta$ is proportional to the width of the distribution.
In fact, comparisons of the obtained localization lengths (e.g., see Fig.~\ref{f:pqrange}) with the prediction from the dispersion relation~\eqref{e:disspec} suggest that the scaling results can not be described by unique values of $\tilde{a}$, $\delta$.  It appears that what we measure in the numerical scaling procedure, is the average of $1/\xi$ for some range of values.  Further work is needed to check and quantify this hypothesis.

\section{Disorder effects in stochastic system}
\label{s:origsto}

In this section we attempt to infer disorder effects in the original stochastic hopping process by mapping-back the results for the equivalent quantum free-fermion model.  In particular, we are interested in disorder-induced changes to the pure steady-state density~\eqref{e:dens}.

\begin{figure}
\begin{center}
\psfrag{w}[Tc][Bc]{$w$}
\psfrag{r}[Bc][Tc]{$\overline{\rho}$}
\psfrag{0}[Tc][Tc]{\scriptsize{0.00}}
\psfrag{0.05}[Tc][Tc]{\scriptsize{0.05}}
\psfrag{0.1}[Tc][Tc]{\scriptsize{0.10}}
\psfrag{0.15}[Tc][Tc]{\scriptsize{0.15}}
\psfrag{0.2}[Tc][Tc]{\scriptsize{0.20}}
\psfrag{0.6665}[Cr][Cr]{\scriptsize{0.6665}}
\psfrag{0.667}[Cr][Cr]{\scriptsize{0.6670}}
\psfrag{0.6675}[Cr][Cr]{\scriptsize{0.6675}}
\psfrag{0.668}[Cr][Cr]{\scriptsize{0.6680}}
\psfrag{0.6685}[Cr][Cr]{\scriptsize{0.6685}}
\psfrag{0.669}[Cr][Cr]{\scriptsize{0.6690}}
\psfrag{"rh05dens.dat"}[Cl][Cl]{\tiny{$p=0.5$, $q=0.5$}}
\psfrag{"rh055dens.dat"}[Cl][Cl]{\tiny{$p=0.55$, $q=0.45$}}
\psfrag{"rh07dens.dat"}[Cl][Cl]{\tiny{$p=0.7$, $q=0.3$}}
\psfrag{"rh10dens.dat"}[Cl][Cl]{\tiny{$p=1.0$, $q=0.0$}}
\psfrag{f(x)}[Cl][Cl]{\tiny{$\rho_0 + b w^2$}}
\includegraphics*[width=1.0\columnwidth]{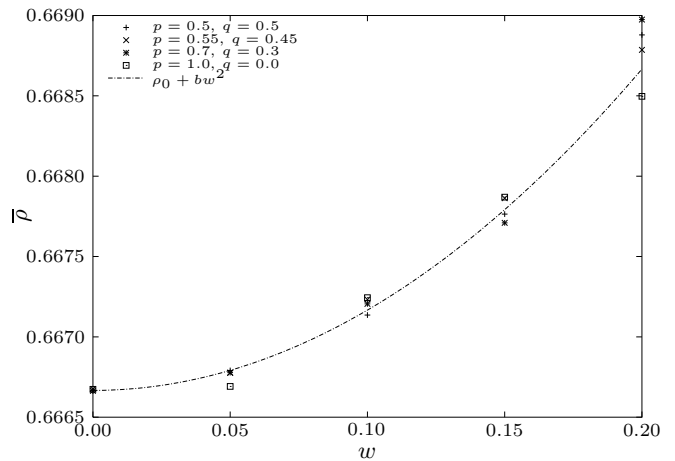}
\caption{Simulation results showing density dependence on width of $\ve$, $\ve'$ distribution for range of $p$, $q$ values (see legend) in a system of size 1000.  $\ve_j$ is drawn from a uniform distribution between $0.8-w$ and $0.8+w$; $\ve'_j$ given by $1-\ve_j$.  Dashed line is a squared fit given by $\overline{\rho}=\rho_0+b w^2$ with $b=0.05$.}
\label{f:storho}
\end{center}
\end{figure}

We argue from the previous section that one can crudely characterize the effects of disorder in the evaporation-deposition rates by the replacements $\ve' \to \ve' \pm i\delta$, $\ve \to \ve \mp i\delta$ ($a_0 \to a_0 \pm i2\delta$) where $\delta$ is related to the width of the disorder distribution.  This gives two different effective Bogoliubov angles (possibly connected to the doubling of fermionic degrees of freedom in the work of Merz and Chalker~\cite{Merz02}).  Stochastic observables are presumably related to the average of these two possibilities but it is not yet clear what is the appropriate function of $\ve$, $\ve'$ to average over.  Motivated by the form of the Bogoliubov transformation [cf.\ Eqs.~\eqref{e:gamrat1} and~\eqref{e:gamrat2}] and by comparisons with simulation, we suggest that the ratio $\ve'/\ve$ is the important quantity and hence define
\begin{equation}
\overline{\left( \frac{\varepsilon'}{\varepsilon} \right)} = \frac{1}{2} \left( \frac{\ve'+i\delta}{\ve-i\delta} + \frac{\ve'-i\delta}{\ve+i\delta} \right).
\end{equation}
This then yields a disorder-averaged density
\begin{equation}
\overline{\rho_l}=\frac{1}{1+\sqrt{\overline{\left(\varepsilon'/\varepsilon \right)}}},
\end{equation}
which is increased from the pure case by an amount proportional to $\delta^2$ (for small $\delta$).  Simulations on the stochastic model (see Fig.~\ref{f:storho}) confirm a density shift proportional to the square of the width of the distribution, i.e., the localization in the disordered free-fermion model does appear to map back to a density shift in the equivalent disordered stochastic system.
Further work is still needed to rigorously establish this picture and to relate the value of $\delta$ both to the details of the disorder distribution and to the numerical localization data.

Similarly, from knowledge of the effective dispersion relation for the disordered case, it should be possible to adapt the pure treatment of~\cite{Grynberg94} for both steady-state dynamic correlation functions and non-steady-state properties.  These quantities typically contain time-dependences of the form $e^{-\text{Re} \lambda_k t}$ 
so any decrease in the effective spectral gap would be expected to lead to a slowing-down of the dynamics (as was found for the ASEP in~\cite{Me04}).

\section{Discussion}
\label{s:ffdiss}

In this paper we have studied the effect of disorder on a model consisting of the partially asymmetric exclusion process combined with dimer evaporation-deposition.  We presented an exact mapping to an equivalent quantum system and then employed a numerical scaling technique on the quantum equations of motion.  This provided a clear demonstration of disorder-induced localization acting in competition with the asymmetric driving.  We discussed tentatively how these localization effects are related to the observed steady-state density shift in the original stochastic system; exact details of the inverse transformation remain to be clarified.  The stochastic system is relatively robust to disorder since particle pairs can be evaporated/deposited along the whole length of the chain. 

There is much scope for further work on this disordered model.  In particular, a more rigorous development of the mapping from the disordered quantum model back to the stochastic system is needed.  This would permit the translation of numerical scaling results for different forms of disorder (including, for example, the case where all rates are disordered and $\ve_j+\ve'_j=p_j+q_j$ is position-dependent) into quantitative statements about the effect on density and correlation functions.  As indicated at the end of Sec.~\ref{s:origsto} we should also like to extend the discussion to dynamics.

Another open question relates to the effect of disorder in the open-boundary case when the evaporation and deposition rates scale with system size, i.e., they are proportional to $1/L$ where $L$ is the number of sites.  In this case the interplay between boundary and bulk effects \emph{is} expected to give a steady-state phase transition in the pure model---compare recent work for a similar dimer model~\cite{Pierobon06} and earlier results for the case with adsorption/desorption of monomers~\cite{Willmann02,Parmeggiani04}.  The addition of disorder to the latter pure monomer model was recently considered in the context of minimal models for intercellular transport~\cite{Pierobon06b}.

\acknowledgments

We would like to thank John Chalker for helpful discussions.  This work was supported by EPSRC under Oxford Condensed Matter Theory Grants GR/R83712/01 and GR/M04426/01 together with Grant No.\ 309261. 

\appendix

\section{Disordered free-fermion scaling}\label{a:ffscale}

Here we present scaling relations for the coupled free-fermion equations of motion~\eqref{e:alpha2} and~\eqref{e:gamma2}:
\begin{widetext}
\begin{align}
(E_{j} - \omega)\Gamma_j &= -p_{j,j-1} \Gamma_{j-1} - q_{j,j+1} \Gamma_{j+1} + \ve_{j,j-1} \gamma_{j-1} -\tilde{\ve}_{j,j+1} \gamma_{j+1}, \label{e:Aalpha2} \\
(\tilde{E}_{j} + \omega)\gamma_j &= -\tilde{p}_{j,j+1} \gamma_{j+1} - \tilde{q}_{j,j-1} \gamma_{j-1} +\ve'_{j,j-1} \Gamma_{j-1} -\tilde{\ve}'_{j,j+1} \Gamma_{j+1}. \label{e:Agamma2}
\end{align}
A $b=2$ decimation of these equations leads to
\begin{align}
(E'_{j} - \omega)\Gamma_j &= -p'_{j,j-2} \Gamma_{j-2} - q'_{j,j+2} \Gamma_{j+2} + (\ve)'_{j,j-2} \gamma_{j-2} -(\tilde{\ve})'_{j,j+2} \gamma_{j+2}, \label{e:Aalpha22} \\
(\tilde{E}'_{j} + \omega)\gamma_j &= -\tilde{p}'_{j,j+2} \gamma_{j+2} - \tilde{q}'_{j,j-2} \gamma_{j-2} +(\ve')'_{j,j-2} \Gamma_{j-2} -(\tilde{\ve'})'_{j,j+2} \Gamma_{j+2}, \label{e:Agamma22}
\end{align}
with rescaled ``energy'' parameters 
\begin{align}
E'_j &= E_j - \frac{p_{j,j-1}q_{j-1,j}}{E_{j-1}-\omega} - \frac{q_{j,j+1}p_{j+1,j}}{E_{j+1}-\omega} + \frac{\ve_{j,j-1}\tilde{\ve'}_{j-1,j}}{\tilde{E}_{j-1}+\omega} + \frac{\tilde{\ve}_{j,j+1}\ve'_{j+1,j}}{\tilde{E}_{j+1}+\omega}, \\
\tilde{E}'_j &= \tilde{E}_j - \frac{\tilde{q}_{j,j-1}\tilde{p}_{j-1,j}}{\tilde{E}_{j-1}+\omega} - \frac{\tilde{p}_{j,j+1}\tilde{q}_{j+1,j}}{\tilde{E}_{j+1}+\omega}  + \frac{\ve'_{j,j-1}\tilde{\ve}_{j-1,j}}{E_{j-1}-\omega} + \frac{\tilde{\ve'}_{j,j+1}\ve_{j+1,j}}{E_{j+1}-\omega},
\end{align}
and the rescaled rates  
\begin{align}
p'_{j,j-2} &= \frac{ -\left( \frac{p_{j,j-1}p_{j-1,j-2}}{E_{j-1}-\omega} + \frac{\ve_{j,j-1}\ve'_{j-1,j-2}}{\tilde{E}_{j-1}+\omega} \right) 
+ \frac{A_{j}}{\tilde{E}'_j + \omega} \left( \frac{\ve'_{j,j-1}p_{j-1,j-2}}{E_{j-1}-\omega} + \frac{\tilde{q}_{j,j-1} \ve'_{j-1,j-2}}{\tilde{E}_{j-1}+\omega} \right) }
{ 1 - \frac{A_{j} B_{j}}{(E'_j-\omega)(\tilde{E}'_j+\omega) } }, \\
q'_{j,j+2} &= \frac{ -\left( \frac{q_{j,j+1}q_{j+1,j+2}}{E_{j+1}-\omega} + \frac{\tilde{\ve}_{j,j+1}\tilde{\ve'}_{j+1,j+2}}{\tilde{E}_{j+1}+\omega} \right) 
- \frac{A_{j}}{\tilde{E}'_j + \omega} \left( \frac{\tilde{\ve'}_{j,j+1}q_{j+1,j+2}}{E_{j+1}-\omega} + \frac{\tilde{p}_{j,j+1} \tilde{\ve'}_{j+1,j+2}}{\tilde{E}_{j+1}+\omega} \right) }
{ 1 - \frac{A_{j} B_{j}}{(E'_j-\omega)(\tilde{E}'_j+\omega) } }, \\
(\ve)'_{j,j-2} &= \frac{ -\left( \frac{p_{j,j-1}\ve_{j-1,j-2}}{E_{j-1}-\omega} + \frac{\ve_{j,j-1}\tilde{q}_{j-1,j-2}}{\tilde{E}_{j-1}+\omega} \right) 
+ \frac{A_{j}}{\tilde{E}'_j + \omega} \left( \frac{\ve'_{j,j-1}\ve_{j-1,j-2}}{E_{j-1}-\omega} + \frac{\tilde{q}_{j,j-1} \tilde{q}_{j-1,j-2}}{\tilde{E}_{j-1}+\omega} \right) }
{ 1 - \frac{A_{j} B_{j}}{(E'_j-\omega)(\tilde{E}'_j+\omega) } }, \\
(\tilde{\ve})'_{j,j+2} &= \frac{ -\left( \frac{q_{j,j+1}\tilde{\ve}_{j+1,j+2}}{E_{j+1}-\omega} + \frac{\tilde{\ve}_{j,j+1}\tilde{p}_{j+1,j+2}}{\tilde{E}_{j+1}+\omega} \right) 
- \frac{A_{j}}{\tilde{E}'_j + \omega} \left( \frac{\tilde{\ve'}_{j,j+1}\tilde{\ve}_{j+1,j+2}}{E_{j+1}-\omega} + \frac{\tilde{p}_{j,j+1} \tilde{p}_{j+1,j+2}}{\tilde{E}_{j+1}+\omega} \right) }
{ 1 - \frac{A_{j} B_{j}}{(E'_j-\omega)(\tilde{E}'_j+\omega) } }, \\
\tilde{p}'_{j,j+2} &= \frac{ -\left( \frac{\tilde{\ve'}_{j,j+1}\tilde{\ve}_{j+1,j+2}}{E_{j+1}-\omega} + \frac{\tilde{p}_{j,j+1} \tilde{p}_{j+1,j+2}}{\tilde{E}_{j+1}+\omega} \right) 
- \frac{B_{j}}{E'_j + \omega} \left( \frac{q_{j,j+1}\tilde{\ve}_{j+1,j+2}}{E_{j+1}-\omega} + \frac{\tilde{\ve}_{j,j+1}\tilde{p}_{j+1,j+2}}{\tilde{E}_{j+1}+\omega} \right) }
{ 1 - \frac{A_{j} B_{j}}{(E'_j-\omega)(\tilde{E}'_j+\omega) } }, \\
\tilde{q}'_{j,j-2} &= \frac{ -\left( \frac{\ve'_{j,j-1}\ve_{j-1,j-2}}{E_{j-1}-\omega} + \frac{\tilde{q}_{j,j-1} \tilde{q}_{j-1,j-2}}{\tilde{E}_{j-1}+\omega} \right) 
+ \frac{B_{j}}{E'_j + \omega} \left( \frac{p_{j,j-1}\ve_{j-1,j-2}}{E_{j-1}-\omega} + \frac{\ve_{j,j-1}\tilde{q}_{j-1,j-2}}{\tilde{E}_{j-1}+\omega} \right) }
{ 1 - \frac{A_{j} B_{j}}{(E'_j-\omega)(\tilde{E}'_j+\omega) } }, \\
(\ve')'_{j,j-2} &= \frac{ -\left( \frac{\ve'_{j,j-1}p_{j-1,j-2}}{E_{j-1}-\omega} + \frac{\tilde{q}_{j,j-1} \ve'_{j-1,j-2}}{\tilde{E}_{j-1}+\omega} \right) 
+ \frac{B_{j}}{E'_j + \omega} \left( \frac{p_{j,j-1}p_{j-1,j-2}}{E_{j-1}-\omega} + \frac{\ve_{j,j-1}\ve'_{j-1,j-2}}{\tilde{E}_{j-1}+\omega} \right) }
{ 1 - \frac{A_{j} B_{j}}{(E'_j-\omega)(\tilde{E}'_j+\omega) } }, \\
(\tilde{\ve'})'_{j,j+2} &= \frac{ -\left( \frac{\tilde{\ve'}_{j,j+1}q_{j+1,j+2}}{E_{j+1}-\omega} + \frac{\tilde{p}_{j,j+1} \tilde{\ve'}_{j+1,j+2}}{\tilde{E}_{j+1}+\omega} \right) 
- \frac{B_{j}}{E'_j + \omega} \left( \frac{q_{j,j+1}q_{j+1,j+2}}{E_{j+1}-\omega} + \frac{\tilde{\ve}_{j,j+1}\tilde{\ve'}_{j+1,j+2}}{\tilde{E}_{j+1}+\omega} \right) }
{ 1 - \frac{A_{j} B_{j}}{(E'_j-\omega)(\tilde{E}'_j+\omega) } }
\end{align}
where 
\begin{align}
A_{j} &= \left( \frac{p_{j,j-1}\tilde{\ve}_{j-1,j}}{E_{j-1}-\omega} -\frac{q_{j,j+1} \ve_{j+1,j}}{E_{j+1}-\omega} - \frac{\ve_{j,j-1}\tilde{p}_{j-1,j}}{\tilde{E}_{j-1}+\omega} + \frac{\tilde{\ve}_{j,j+1}\tilde{q}_{j+1,j}}{\tilde{E}_{j+1}+\omega} \right), \\
B_{j} &= \left(- \frac{\ve'_{j,j-1}q_{j-1,j}}{E_{j-1}-\omega} + \frac{\tilde{\ve'}_{j,j+1} p_{j+1,j}}{E_{j+1}-\omega} + \frac{\tilde{q}_{j,j-1}\tilde{\ve'}_{j-1,j}}{\tilde{E}_{j-1}+\omega} - \frac{\tilde{p}_{j,j+1}\ve'_{j+1,j}}{\tilde{E}_{j+1}+\omega} \right).
\end{align}
Note the difference in meaning between $\ve'$ (the evaporation rate) and $(\ve)'$ (the scaled deposition rate).  
\end{widetext}

\bibliographystyle{apsrev}
\bibliography{/home/usersHR/harris/allref}

\end{document}